\documentclass[12pt]{article}
\usepackage{amssymb} 
\makeatletter
\newcommand{\rddots}{\mathinner{\mkern1mu\raise\p@\vbox{\kern7\p@\hbox{.}}%
\mkern2mu\raise4\p@\hbox{.}\mkern2mu\raise7\p@\hbox{.}\mkern1mu}}
\@addtoreset{equation}{section}

\makeatother

\newcommand{\Hol}{{\mathcal H}}
\newcommand{\Cl}{{\rm Cl}}

\newcommand{\R}{{\rm R}}
\newcommand{\Hq}{{\rm H}}

\newcommand{\E}{{\rm E}}
\newcommand{\SO}{{\rm SO}}
\newcommand{\SL}{{\rm SL}}
\newcommand{\GL}{{\rm GL}}
\newcommand{\Spin}{{\rm Spin}}
\newcommand{\ISO}{{\rm ISO}}
\newcommand{\SU}{{\rm SU}}
\newcommand{\Sp}{{\rm Sp}}

\newcommand{\D}{{\mathcal D}}
\newcommand{\U}{{\rm U}}

\newcommand{\beq}{\begin{equation}}
\newcommand{\eeq}[1]{\label{#1}\end{equation}}
\newcommand{\bea}{\begin{eqnarray}}
\newcommand{\eea}[1]{\label{#1}\end{eqnarray}}

\newcommand{\fft}[2]{{\frac{#1}{#2}}}
\newcommand{\ft}[2]{{\textstyle{\frac{#1}{#2}}}}
\ifx\Bbb\undefined
\else
\renewcommand{\R}{{\mathbb R}}
\renewcommand{\Hq}{{\mathbb H}}

\renewcommand{\E}{{\mathbb E}}
\fi
\ifx\ltimes\undefined
\newcommand{\ltimes}{{\kern3pt\hbox{\vrule width 0.4pt height 5.30pt depth
.0pt}\kern-1.76pt\times\kern1pt}}
\fi
\begin{document}

\begin{titlepage}
\begin{flushright}
MCTP-03-58\\
hep-th/0312165
\end{flushright}

\vspace{15pt}

\begin{center}
{\large\bf Generalized holonomy of M-theory vacua\footnote{Research
supported in part by DOE Grant DE-FG02-95ER40899.}}

\vspace{15pt}

{A.~Batrachenko\footnote{abat@umich.edu},
M.~J.~Duff\footnote{mduff@umich.edu},
James T.~Liu\footnote{jimliu@umich.edu} and
W.~Y.~Wen\footnote{wenw@umich.edu}}

\vspace{7pt}

{\it Michigan Center for Theoretical Physics\\
Randall Laboratory, Department of Physics, University of Michigan\\
Ann Arbor, MI 48109--1120, USA}

\end{center}

\begin{abstract}

The number of M-theory vacuum supersymmetries, $0\leq n\leq 32$, is
given by the number of singlets appearing in the decomposition of the
$32$ of $\SL(32,\R)$ under $\Hol \subset \SL(32,\R)$ where $\Hol$
is the holonomy group of the generalized connection which incorporates
non-vanishing $4$-form.  Here we compute this generalized holonomy for 
the $n=16$ examples of the M2-brane, M5-brane, M-wave, M-monopole,  
for a variety of their $n=8$ intersections and also for the $n >16$ pp waves.

\end{abstract}

\end{titlepage}


\section{Introduction}
\label{Introduction}

The equations of M-theory display the maximum number of supersymmetries
$N=32$, and so $n$, the number of supersymmetries preserved by a
particular vacuum, must be some integer $0\leq n\leq 32$.  In vacua
with vanishing 4-form $F_{(4)}$, it is well known that $n$ is given by the
number of singlets appearing in the decomposition of the 32 of $\SO(1,10)$
under ${H} \subset \SO(1,10)$ where $H$ is the holonomy group of the
usual Riemannian connection.
\begin{equation}
D_{M}=\partial_{M}+\frac{1}{4}\omega_{M}{}^{AB}\Gamma_{AB}.
\end{equation}
Here $\Gamma_{A}$ are the $\SO(1,10)$ Dirac matrices and
$\Gamma_{AB}=\Gamma_{[A}\Gamma_{B]}$. This connection can account for vacua
with $n=0$, 1, 2, 3, 4, 6, 8, 16, 32.

Vacua with non-vanishing $F_{(4)}$ allow more exotic fractions of
supersymmetry, including $16<n<32$.  Here, however, it is necessary
to generalize the notion of holonomy to accommodate the generalized
connection that results from a non-vanishing $F_{(4)}$.
\begin{equation}
\label{covariant}
{\cal D}_{M}=D_{M}
-\frac{1}{288}(\Gamma_M{}^{NPQR}-8\delta_M^N\Gamma^{PQR})F_{NPQR}.
\label{supercovariant}
\end{equation}
As discussed in a previous paper \cite{Duff:2003ec}, the number of
M-theory vacuum supersymmetries is now given by the number of singlets
appearing in the decomposition of the $32$ of ${\cal G}$ under $\Hol
\subset \cal G$ where $\cal G$ is the generalized structure group and 
$\Hol$ is the generalized holonomy group.  Discussions of
generalized holonomy may also be found in \cite{Duffstelle,Duff}.

In subsequent papers by Hull \cite{Hull:2003mf} and Papadopoulos
and Tsimpis \cite{Papadopoulos:2003pf} it was shown that $\cal G$ may
be as large as $\SL(32,\R)$ and that an M-theory vacuum admits precisely $n$
Killing spinors iff
\begin{equation}
\SL(31-n,\R) \ltimes (n+1)\R^{(31-n)} \supseteq\kern-12pt/\kern8pt {\Hol}
\subseteq  \SL(32-n,\R)\ltimes n\R^{(32-n)},
\label{n}
\end{equation}
{\it i.e.} the generalized holonomy is contained in $\SL(32-n,\R)\ltimes
n\R^{(32-n)}$ but is not contained in $\SL(31-n,\R) \ltimes
(n+1)\R^{(31-n)}$.

In this paper we compute this generalized holonomy for the $n=16$ examples
of the M2-brane, M5-brane, M-wave (MW) and the M-monopole (MK), for a variety of their 
$n=8$ intersections: M2/MW, M5/MW, M2/M5, MW/MK and also for the 
$n>16$ ppwaves. We begin with a review of generalized holonomy in section 2.  Then
we turn to $n=16$ and $n=8$ solutions in sections 3 and 4.  Since pp-waves
with exotic fractions of supersymmetry involve a slightly different analysis,
they are covered in section 5.  Finally, we conclude with some comments in
section 6.

\section{Holonomy and supersymmetry}
\label{Riemannian}

The number of supersymmetries preserved by an M-theory background
depends on the number of covariantly constant spinors, %
\begin{equation}
{\cal D}_{M}\epsilon=0,
\label{integrability1}
\end{equation}
called {\it Killing} spinors.  It is the presence of the terms involving
the 4-form $F_{(4)}$ in (\ref{covariant}) that makes this counting difficult.
So let us first examine the simpler vacua for which $F_{(4)}$ vanishes.
Killing spinors then satisfy the integrability condition
\begin{equation}
[{D}_{M}, {D}_{N}] \epsilon=\frac{1}{4}R_{MN}{}^{AB}\Gamma_{AB}\epsilon=0,
\label{integrability2}
\end{equation}
where $R_{MN}{}^{AB}$ is the Riemann tensor.  The subgroup of
$\Spin(10,1)$ generated by this linear combination of $\Spin(10,1)$ generators
$\Gamma_{AB}$ corresponds to the ${\it holonomy}$ group ${H}$ of the
connection $\omega_{M}$.  We note that the same information is
contained in the first order Killing spinor equation (\ref{integrability1})
and second-order integrability condition (\ref{integrability2}).  One
implies the other, at least locally.  The number of
supersymmetries, $n$, is then given by the number of singlets
appearing in the decomposition of the $32$ of $\Spin(10,1)$ under
${H}$.  In Euclidean signature, connections satisfying
(\ref{integrability2}) are automatically Ricci-flat and hence solve
the field equations when $F_{(4)}=0$.  In Lorentzian signature, however,
they need only be Ricci-null so Ricci-flatness has to be imposed as an
extra condition.  In Euclidean signature, the holonomy groups have
been classified \cite{Berger}.  In Lorentzian signature, much less is
known but the question of which subgroups ${H}$ of $\Spin(10,1)$ leave
a spinor invariant has been answered \cite{Bryant}.  There are two sequences
according as the Killing vector
$v_{A}=\overline{\epsilon}\,\Gamma_{A}\epsilon$ is timelike or null.
Since $v^{2} \leq 0$, the spacelike $v_A$ case does not arise.  The
timelike $v_A$ case corresponds to static vacua, where ${H} \subset
\Spin(10) \subset \Spin(10,1)$ while the null case to non-static vacua
where ${H} \subset \ISO(9) \subset \Spin(10,1)$.  It is then possible to
determine the possible $n$-values and one finds $n=2$, 4, 6, 8, 16, 32
for static vacua, and $n=1$ 2, 3, 4, 8, 16, 32 for non-static vacua
\cite{Acharya:1998yv,Acharya:1998st,Fig}.

\subsection{Generalized holonomy}
\label{Generalized}

In general we want to include vacua with $F_{(4)}\neq 0$.  Such vacua
are physically interesting for a variety of reasons. In particular, they
typically have fewer moduli than their zero $F_{(4)}$ counterparts
\cite{Duffnilssonpopecomp}. Now, however, we face the problem that the
connection in (\ref{covariant}) is no longer the spin connection to which
the bulk of the mathematical literature on holonomy groups is devoted.
In addition to the $\Spin (10,1)$ generators $\Gamma_{AB}$, it is apparent
from (\ref{supercovariant}) that there are terms involving $\Gamma_{ABC}$
and $\Gamma_{ABCDE}$.  In fact, the generalized connection takes its values
in $\SL(32,\R)$.  Note, however, that some generators are missing from the
covariant derivative.  Denoting the antisymmetric product of $k$ Dirac
matrices by $\Gamma^{(k)}$, the complete set of $\SL(32,\R)$ generators
involve $\{\Gamma^{(1)},\Gamma^{(2)},
\Gamma^{(3)},\Gamma^{(4)},\Gamma^{(5)}\}$ whereas only $\{\Gamma^{(2)},
\Gamma^{(3)},\Gamma^{(5)}\}$ appear in the covariant derivative.
Another way in which generalized holonomy differs from the Riemannian
case is that, although the vanishing of the covariant derivative of the
spinor implies the vanishing of the commutator, the converse is not
true, as discussed below in section \ref{sec:intr}.

This generalized connection can preserve exotic fractions of
supersymmetry forbidden by the Riemannian connection.  For example,
M-branes at angles \cite{Ohta} include $n$=5, 11-dimensional pp-waves
\cite{Michelson,Cvetic:2002si,Gauntletthull2,Bena:2002kq} include
$n=18$, 20, 22, 24, 26, squashed $N(1,1)$ spaces \cite{Page} and
M5-branes in a pp-wave background \cite{Singh} include $n=12$ and
G\"{o}del universes \cite{Gauntlett:2002nw, Harmark} include $n=14$, 18,
20, 22, 24.  However, we can attempt to quantify this in terms of
generalized holonomy groups%
\footnote{In this paper we focus on $D=11$ but similar generalized holonomy
can be invoked to count $n$ in Type IIB vacua \cite{Papadopoulos}, which
include pp-waves with $n=28$ \cite{Bena:2002kq}.}.
Generalized holonomy means that one can assign a holonomy ${\cal H}
\subset {\cal G}$ to the generalized connection appearing in the
supercovariant derivative ${\cal D}$ where ${\cal G}$ is the generalized
structure group.  The number of unbroken supersymmetries is then given by
the number of ${\cal H}$ singlets appearing in the decomposition of the
32 dimensional representation of ${\cal G}$ under $\Hol \subset {\cal G}$.

For generic backgrounds we require that ${\cal G}$ be the full $\SL(32,\R)$
while for special backgrounds smaller ${\cal G}$ are sufficient
\cite{Hull:2003mf}.  To see this, let us write the supercovariant
derivative as
\begin{equation}
{\cal D}_{M}=\hat D_{M}+X_{M},
\label{split2}
\end{equation}
for some other connection $\hat D_{M}$ and some covariant $32 \times 32$
matrix $X_{M}$. If we now specialize to backgrounds satisfying
\begin{equation}
X_{M}\epsilon=0,
\label{X}
\end{equation}
then the relevant structure group is ${\hat G} \subseteq {\cal G}$.

Consider, for example, for the connection ${\hat D}$ arising in dimensional
reduction of $D=11$ supergravity. One can show \cite{Duff:2003ec} that the
lower dimensional gravitino transformation may be written
\begin{equation}
\delta\psi_\mu=\hat D_\mu\epsilon,
\label{hat}
\end{equation}
in terms of a covariant derivative
\begin{equation}
\hat D_\mu=\partial_\mu+\omega_\mu{}^{\alpha\beta}\gamma_{\alpha\beta}
+Q_\mu{}^{ab}\Gamma_{ab}+\ft1{3!}e^{ia}e^{jb}e^{kc}\partial_\mu\phi_{ijk}
\Gamma_{abc}.
\label{hat1}
\end{equation}
Here $\gamma_\alpha$ are $\SO(d-1,1)$ Dirac matrices, while $\Gamma_a$
are $\SO(11-d)$ Dirac matrices. In the above, the lower dimensional quantities
are related to their $D=11$ counterparts $(E_M{}^A,\Psi_M^{(11)},A_{MNP})$
through
\begin{eqnarray}
&&ds_{(11)}^2=\Delta^{-\fft1{d-2}}ds_4^2+g_{ij}dy^idy^j,\nonumber\\
&&\psi_\mu=\Delta^{\fft1{4(d-2)}}\left(\Psi^{(11)}_\mu+\fft1{d-2}\gamma_\mu
\Gamma^i\Psi^{(11)}_i\right),\qquad
\lambda_i=\Delta^{\fft1{4(d-2)}}\Psi^{(11)}_i,\nonumber\\
&&\epsilon=\Delta^{\fft1{4(d-2)}}\epsilon^{(11)},\nonumber\\
&&Q^{ab}_\mu=e^{i[a}\partial_\mu e_i{}^{b]},\qquad
P_{\mu\,ij}=e_{(i}^a\partial_\mu e_{j)\,a},\qquad
\phi_{ijk}=A_{ijk}.
\end{eqnarray}
The condition (\ref{X}) is just $\delta \lambda_{i}=0$ where
$\lambda_{i}$ are the dilatinos of the dimensionally reduced theory.
In this case, the generalized holonomy is given by ${\hat H} \subseteq
{\hat G}$ where the various $\hat G$ arising in spacelike, null and
timelike compactifications are tabulated in \cite{Duff:2003ec} for
different numbers of the compactified dimensions. These smaller 
structure groups are also the ones appropriate to more general 
Kaluza-Klein compactifications of the product manifold type,
{\it i.e.} without a warp factor \cite{Hull:2003mf}. 

This is probably a good time to say a few words about the difference 
between generalized holonomy and the hidden symmetries conjecture 
which were both discussed in \cite{Duff:2003ec}. There it was argued that the equations 
of M-theory possess previously unidentified hidden spacetime (timelike and null)
symmetries in addition to the well-known hidden internal (spacelike)
symmetries. They take the form ${\cal
G}=\SO(d-1,1) \times G(spacelike)$, ${\cal G}= \ISO(d-1) \times
G(null)$ and ${\cal G}=\SO(d) \times G(timelike)$ with $1\leq d<11$.
For example, $G(spacelike)=\SO(16)$, $G(null)=[\SU(8) \times
\U(1)]\ltimes \R^{56}$ and $G(timelike)=\SO^*(16)$ when $d=3$.  The
nomenclature derives from the fact that they coincide with
the hidden symmetry groups that appear in the spacelike, null and timelike 
dimensional reductions of the theory.  However, they were proposed as
background-independent symmetries of the full unreduced and untruncated $D=11$ 
equations of motion, not merely their dimensional reduction.

For $d \geq 3$, these coincide with
the generalized structure groups $\hat G$ discussed above that appear in 
the dimensionally  reduced covariant derivative. A more speculative idea is 
that there exists a yet-to-be-discovered version of $D=11$ supergravity or 
$M$-theory that displays even bigger hidden symmetries corresponding to 
${\hat G}$ with $d<3$ \cite{Duff:2003ec} which could be as large as 
$\SL(32,\R)$ \cite{Hull:2003mf}. 

To avoid possible confusion, we emphasize here that the notion of generalized 
holonomy ${\cal H} \subset \SL(32,\R)$ is valid whether or not these hidden 
symmetry conjectures turn out to be correct, and the misleading phrase 
`generalized holonomy conjecture' should now be abandoned. This 
highlights another difference between generalized and Riemannian  
holonomy, $\cal H $ need not be a symmetry of the theory, whereas 
$H\subset \SO(1,10)$ always is. 

Note also that a recent paper \cite{Keurentjes} calls into question both 
generalized holonomy and the hidden symmetries conjecture in the 
presence of fermions because some of the symmetries do not admit spinor 
representations. While acknowledging that such global considerations are 
important, we do not take the same pessimistic attitude since the fermions 
do not generally transform as spinors. In the $d=3$ spacelike case, 
for example, the gravitino transforms as a {\it vector} 16 of $\SO(16)$. 

\subsection{Integrability conditions}
\label{sec:intr}

Yet another way in which generalized holonomy differs from Riemannian
holonomy is that, although the vanishing of the covariant derivative
implies the vanishing of the commutator, the converse is not true.
Consequently, the second order integrability condition alone may be a
misleading guide to the generalized holonomy group $\cal H$.

To illustrate this, we consider Freund-Rubin \cite{Freundrubin} vacua with
$F_{(4)}$ given by
\begin{equation}
F_{\mu\nu\rho\sigma}=3m\epsilon_{\mu\nu\rho\sigma},
\end{equation}
where $\mu=0,1,2,3$ and $m$ is a constant with the dimensions of mass.
This leads to an $AdS_{4} \times X^{7}$ geometry.  For such a product
manifold , the supercovariant derivative splits as
\begin{equation}
{\cal D}_{\mu}= D_{\mu}+m\gamma_{\mu}\gamma_{5}
\end{equation}
and
\begin{equation}
{\cal D}_{m}= D_{m}-\ft{1}{2}m\Gamma_{m},
\label{covariant1}
\end{equation}
and the Killing spinor equations reduce to
\begin{equation}
{\cal D}_{\mu}\epsilon(x) = 0
\end{equation}
and
\begin{equation}
{\cal D}_{m}\eta(y)= 0.
\label{Killing}
\end{equation}
Here $\epsilon(x)$ is a 4-component spinor and $\eta(y)$ is an
8-component spinor, transforming with Dirac matrices $\gamma_\mu$ and
$\Gamma_m$ respectively.  The first equation is satisfied automatically
with our choice of $AdS_{4}$ spacetime and hence the number of $D=4$
supersymmetries, $0\leq N \leq 8$, devolves upon the number of
Killing spinors on $X^{7}$ \cite{Wittenkaluza}.  They satisfy the integrability condition
\begin{equation}
[{\cal D}_{m}, {\cal D}_{n}] \eta=
-\frac{1}{4}C_{mn}{}^{ab}\Gamma_{ab}\eta=0,
\label{integrability}
\end{equation}
where $C_{mn}{}^{ab}$ is the Weyl tensor. Owing to this generalized
connection, vacua with $m\neq 0$ present subtleties and novelties not
present in the $m=0$ case
\cite{vanNwarner}, for example the phenomenon of
{\it skew-whiffing} \cite{Duffnilssonpopesuper,Duffnilssonpopekaluza}.
For each Freund-Rubin compactification, one may obtain another by
reversing the orientation of $X^{7}$.  The two may be distinguished by
the labels {\it left} and {\it right}.  An equivalent way to obtain
such vacua is to keep the orientation fixed but to make the
replacement $m\rightarrow -m$ thus reversing the sign of $F_{4}$.  So
the covariant derivative (\ref{covariant1}), and hence the condition
for a Killing spinor, changes but the integrability condition
(\ref{integrability}) remains the same.  With the exception of the
round $S^{7}$, where both orientations give $N=8$, at most one
orientation can have $N \geq 0$.  This is the {\it skew-whiffing
theorem}.  (Note, however, that skew-whiffed vacua are automatically
stable at the classical level since skew-whiffing affects only the
spin $3/2$, $1/2$ and $0^{-}$ towers in the Kaluza-Klein spectrum,
whereas the criterion for classical stability involves only the
$0^{+}$ tower \cite{DNP,Duffnilssonpopekaluza}.)

The squashed $S^{7}$ provides a non-trivial example
\cite{Awadaduffpope,Duffnilssonpopesuper}: the left squashed
$S^{7}$ has $N=1$ but the right squashed $S^{7}$ has $N=0$.  Other
examples are provided by the left squashed $N(1,1)$ spaces \cite{Page},
one of which has $N=3$ and the other $N=1$, while the right squashed
counterparts both have $N=0$.  (Note, incidentally, that $N=3$ {\it i.e.}
$n=12$ can never arise in the Riemannian case.)

All this presents a dilemma.  If the Killing spinor
condition changes but the integrability condition does not, how does
one give a holonomic interpretation to the different supersymmetries?
We note that in (\ref{covariant1}), the $SO(7)$ generators $\Gamma_{ab}$,
augmented by
presence of $\Gamma_{a}$, together close on $SO(8)$ \cite{Castellani}.
Hence the generalized holonomy group satisfies ${\cal H}\subset SO(8)$.
We now ask how the $8$ of $SO(8)$ decomposes under ${\cal H}$.  In the
case of the left squashed $S^{7}$, ${\cal H}= SO(7)^{-}$, and $N=1$, but for the right squashed $S^{7}$, ${\cal
H}= SO(7)^{+}$, $8 \rightarrow 8$ and $N=0$.  From the integrability
condition alone, however, we would have concluded naively that ${\cal
H}=G_{2}\subset SO(7)$  for which $8\rightarrow 1+7$ and hence that both 
orientations give $N=1$.

\subsection{Higher order corrections}

Another context in which generalized holonomy may prove important is
that of higher loop corrections to the M-theory Killing spinor equations with
or without the presence of non-vanishing $F_{(4)}$.  As discussed in
\cite{Lu:2003ze}, higher loops yield non-Riemannian corrections to
the supercovariant derivative, even for vacua for which $F_{(4)}=0$,
thus rendering the Berger classification inapplicable. Although the Killing
spinor equation receives higher order corrections, so does the metric,
ensuring, for example, that $H=G_{2}$ Riemannian holonomy 7-manifolds still
yield $N=1$ in $D=4$ when the non-Riemannian corrections are taken
into account.  This would require a generalized holonomy ${\cal H}$
for which the decomposition $8 \rightarrow 1+7$ continues to hold.

\section{Generalized holonomy for $n=16$}

We now turn to a generalized holonomy analysis of some basic supergravity
solutions.  Starting with the maximally supersymmetric backgrounds ($n=32$),
namely $\E^{1,10}$, AdS$_7\times S^4$, AdS$_4\times S^7$ and Hpp, it should
be clear that they all have trivial generalized holonomy, in accord with
(\ref{n}).  However, only flat space may be described by (trivial)
Riemannian holonomy.

Somewhat more interesting to consider are the four basic objects of M-theory
preserving half of the supersymmetries (corresponding to $n=16$).  These are
the M5-brane, M2-brane, M-wave (MW) and the Kaluza-Klein monopole (MK).  The
latter two have $F_{(4)}=0$ and may be categorized using ordinary Riemannian
holonomy, with $H\subset\SO(10,1)$.  We now look at these in turn.

\subsection{The M5-brane}
\label{sec:m5sec}

The familiar supergravity M5-brane solution \cite{Gueven} may be written
in isotropic coordinates as
\begin{eqnarray}
&&ds^2=H_5^{-1/3}dx_\mu^2+H_5^{2/3}d\vec y\,^2,\nonumber\\
&&F_{ijkl}=\epsilon_{ijklm}\partial_mH_5,
\label{eq:m5soln}
\end{eqnarray}
where $H_5(\vec y\,)$ is harmonic in the six-dimensional transverse space
spanned by $\{y^i\}$, and $\epsilon_{ijklm}=\pm1$.  While the transverse
space only needs to be Ricci flat, we take it to be $\E^5$, so as
not to further break the supersymmetry.

A simple computation of the generalized covariant derivative on this
background yields
\begin{eqnarray}
\D_\mu&=&\partial_\mu-\ft16\Gamma_{\bar\mu}{}^{\bar i}P_5^+H^{-3/2}
\partial_iH,\nonumber\\
\D_i&=&\partial_i+\ft13\Gamma_{\bar i}{}^{\bar j}P_5^+\partial_j
\ln H-\ft12\Gamma^{(5)}\partial_i\ln H.
\end{eqnarray}
Here, $P_5^\pm=\fft12(1\pm\Gamma^{(5)})$ is the standard 1/2-BPS projection
for the M5-brane, where $\Gamma^{(5)}=\fft1{5!}\epsilon_{ijklm}\Gamma^{\bar i
\bar j\bar k\bar l\bar m}$.  All quantities with bars indicate tangent space
indices.  To obtain the generalized holonomy of the
M5-brane, we examine the commutator of covariant derivatives.  Defining
\begin{equation}
M_{MN}=[\D_M,\D_N],
\label{eq:comm}
\end{equation}
we find that $M_{\mu\nu}=0$, so that the holonomy is trivial in
the longitudinal directions along the brane.  On the other hand, the
transverse and mixed commutators are given by
\begin{eqnarray}
M_{ij}&=&-\ft29\Gamma_{\bar i\bar j}P_5^+(\partial_k\ln H)^2
+\ft23\Gamma_{[\bar j}{}^{\bar k}P_5^+(\partial_i\partial^k\ln H
-\ft23\partial_{i]}\ln H\partial^k\ln H),\nonumber\\
M_{\mu i}&=&H^{-1/2}[\ft16\Gamma_{\bar\mu\bar j}P_5^+(\partial_i
\partial^j\ln H-\ft23\partial_i\ln H\partial_j\ln H)+\ft1{18}\Gamma_{\bar\mu
\bar i}P_5^+(\partial_j\ln H)^2].
\end{eqnarray}

We first examine the transverse holonomy.  Independent of the form of the
harmonic function, $H_5$, we see that the only combination of Dirac matrices
showing up in $M_{ij}$ are given by $\Gamma_{\bar i\bar j}P_5^+$.  Defining
a set of Hermitian generators $T_{ij}=-\ft{i}2\Gamma_{\bar i\bar j}P_5^+$, it
is easily seen that they generate the $\SO(5)$ algebra
\begin{equation}
[T_{ij},T_{kl}]= i(\delta_{ik}T_{jl}-\delta_{il}T_{jk}-\delta_{jk}T_{il}
+\delta_{jl}T_{ik}).
\end{equation}
As a result, the transverse holonomy is simply $\SO(5)_+$, where the $+$
refers to the sign of the M5-projection.

Turning next to the mixed commutator, $M_{\mu i}$, we see that it introduces
an additional set of Dirac matrices, $K_{\mu i}=\Gamma_{\bar\mu\bar i}P_5^+$.
Since $\Gamma_{\bar\mu}P_5^+=P_5^-\Gamma_{\bar\mu}$, it is clear that the
$K_{\mu i}$ generators commute among themselves.  On the other hand,
commuting $K_{\mu i}$ with the $\SO(5)_+$ generators $T_{ij}$ yield the
additional combinations $K_\mu=\Gamma_{\bar\mu}P_5^+$ and $K_{\mu ij}
=\Gamma_{\bar\mu\bar i \bar j}P_5^+$.  Picking a set of Cartan generators
$T_{12}$ and $T_{34}$ for $\SO(5)_+$, we may see that the complete set
$\{K_\mu,K_{\mu i},K_{\mu ij}\}$ has weights $\pm1/2$.  As a result, they
transform as a set of 4-dimensional spinor representations of $\SO(5)_+$.
We conclude that the generalized holonomy of the M5-brane is
\begin{equation}
\Hol_{\rm M5}=\SO(5)_+\ltimes6\R^{4(4)}.
\label{eq:m5hol}
\end{equation}

\subsection{The M2-brane}

Turning next to the M2-brane, its supergravity solution may be written
as \cite{Duffstelle}
\begin{eqnarray}
&&ds^2=H_2^{-2/3}dx_\mu^2+H_2^{1/3}d\vec y\,^2,\nonumber\\
&&F_{\mu\nu\rho i}=\epsilon_{\mu\nu\rho}\partial_i\fft1{H_2}.
\label{eq:m2soln}
\end{eqnarray}
A similar examination of the commutator of generalized covariant derivatives,
(\ref{eq:comm}), for this solution indicates the presence of both
compact generators $T_{ij}=-\fft{i}2\Gamma_{\bar i\bar j}P_2^+$ and
non-compact ones $K_{\mu i}=\Gamma_{\bar\mu\bar i}P_2^+$.  Here, $P_2^\pm
=\fft12(1\pm\Gamma^{(2)})$ where $\Gamma^{(2)}=\fft1{3!}\epsilon_{\mu\nu
\rho}\Gamma^{\bar\mu\bar\nu\bar\rho}$ is the M2-brane projection.
Furthermore, the coordinates on (\ref{eq:m2soln}) correspond to a $3/8$
longitudinal/transverse split.  Hence the transverse holonomy in this
case is $\SO(8)_+$.

To obtain the generalized holonomy group $\Hol_{M2}$, we must first close
the algebra formed by $T_{ij}$ and $K_{\mu i}$.  Upon doing so, we find the
additional generators $K_{\mu ijk}=\Gamma_{\bar\mu\bar i\bar j\bar k}P_2^+$.
As in the M5 case, we may see that the set $\{K_{\mu i},K_{\mu ijk}\}$ form
eight-dimensional representations of $\SO(8)_+$.  However, some care must
be taken in identifying these representations as the $8_v$, $8_s$ or
$8_c$ (up to an overall automorphism due to triality).

Since it is instructive for the later intersecting brane examples, we will
demonstrate a simple method for investigating the generalized holonomy of
this solution.  Based on the $3/8$ split, we may make an explicit
decomposition of the 11-dimensional (real) Dirac matrices as follows
\begin{eqnarray}
\Gamma^0&=&1\times i\sigma^2\times 1,\nonumber\\
\Gamma^1&=&1\times \sigma^1\times 1,\nonumber\\
\Gamma^2&=&1\times \sigma^3\times \sigma^3,\nonumber\\
\Gamma^3&=&1\times\sigma^3\times\sigma^1,\nonumber\\
\Gamma^a&=&\gamma^a\times\sigma^3\times\sigma^2.
\label{eq:m2dirac}
\end{eqnarray}
Here, the eight-dimensional transverse space is split into $7+1$,
with $\gamma^a$ a set of purely imaginary $8\times8$ seven-dimensional
Dirac matrices.  Since $\Gamma^{(2)}\equiv\Gamma^{012}=1\times1\times
\sigma^3$, the M2-brane projection is simply
\begin{equation}
P_2^+=1\times1\times\pmatrix{1&0\cr0&0}.
\label{eq:m2proj}
\end{equation}

The explicit $\SO(8)_+$ generators then have the form
\begin{equation}
T_{ij}\quad\longleftrightarrow\quad\{-\ft{i}2\gamma^{ab},\ft12\gamma^a\}
\times1\times\pmatrix{1&0\cr0&0},
\label{eq:m2sl}
\end{equation}
which highlights the embedding of $\SO(7)\subset\SO(8)_+$.  The complete
set of mixed generators may be written concisely as
\begin{equation}
\{K_{\mu i},K_{\mu ijk}\}\quad\longleftrightarrow\quad \Cl(0,7)_+
\times\{1,\sigma^1,i\sigma^2\}\times\pmatrix{0&0\cr1&0},
\label{eq:m2r}
\end{equation}
where $\Cl(p,q)$ is the real Clifford algebra with signature given by
$p$ positive and $q$ negative eigenvalues.  In this case, $\Cl(0,7)_+$
is generated by the Dirac matrices $i\gamma^a$, and is isomorphic to
$\GL(8,\R)$.

Examination of (\ref{eq:m2sl}) and (\ref{eq:m2r}) demonstrates that the
M2 holonomy generators have the schematic form
\begin{equation}
\pmatrix{\SO(8)_+\times1&0\cr\R(8,8)\times\{1,\sigma^1,i\sigma^2\}&0}
\subset\pmatrix{\SL(16,\R)&0\cr\R(16,16)&0},
\label{eq:m2holgen}
\end{equation}
as appropriate to a solution with $n=16$.  This shows that the M2 generalized
holonomy is given by
\begin{equation}
\Hol_{\rm M2}=\SO(8)_+\ltimes12\R^{2(8_s)}.
\label{eq:m2hol}
\end{equation}
This corrects a result obtained in \cite{Duffstelle} where it was
claimed that the generalized holonomy is simply $\Hol_{\rm M2}={\hat 
H}_{\rm M2}=\SO(8)_+$ which also yields $n=16$.

\subsection{The M-wave}

We now turn to the pure geometry solutions.  The wave (MW) is given
by \cite{Hull}
\begin{equation}
ds^2=2\,dx^+\,dx^-+K\,dx^{+\,2}+d\vec y\,^2,
\label{eq:mwsoln}
\end{equation}
where $K(\vec y\,)$ is harmonic on the nine-dimensional Euclidean transverse
space $\E^9$.  In a vielbein basis $e^+=dx^+$, $e^-=dx^-+\fft12K\,dx^+$,
$e^i=dy^i$, the only non-vanishing component of the spin connection is given
by $\omega^{+i}=\fft12\partial_iK\,e^+$.  Thus the gravitational covariant
derivative acting on $\epsilon$ is given by
\begin{equation}
D_+=\partial_++\ft14\partial_iK\Gamma_-\Gamma_i,\qquad
D_-=\partial_-,\qquad
D_i=\partial_i.
\label{eq:ppcder}
\end{equation}
Note that the metric is given by $ds^2=2e^+e^-+e^ie^i$, so that light
cone indices are raised and lowered as, {\it e.g.}, $\Gamma_-=\Gamma^+$
in tangent space.

The only non-vanishing commutator of covariant derivatives is given by
\begin{equation}
M_{+i} = -\ft14\partial_i\partial_jK\Gamma_-\Gamma_i,
\label{eq:commw}
\end{equation}
so we may identify the generalized holonomy generators as $T^i=
\Gamma_-\Gamma_i$.  Since $\Gamma_-^2=0$, these nine generators are
mutually commuting, and the MW generalized holonomy is
\begin{equation}
\Hol_{\rm MW}=\R^9.
\end{equation}
In addition to being a subgroup of $\SL(16,\R)\ltimes16\R^{16}$, this may
also be viewed as a subgroup of $\ISO(9)$ appropriate to backgrounds with
a null Killing vector.
We will return to waves in section~\ref{sec:ppwave}, where we turn on
$F_{(4)}$ and consider the generalized holonomy of pp-waves preserving
exotic fractions of supersymmetry.

\subsection{The M-monopole}

The final basic M-theory object we consider is the Kaluza-Klein monopole,
which is given by the Euclidean Taub-NUT solution \cite{Han}
\begin{equation}
ds^2=dx_\mu^2+H(dr^2+r^2d\Omega_2^2)+H^{-1}(dz-q\cos\theta\,d\phi)^2,
\end{equation}
where $d\Omega_2^2=d\theta^2+\sin^2\theta\,d\phi^2$ and $H=1+q/r$.  As
is well known, this space is Ricci flat and hyper-K\"ahler, and so has
$\Sp(1)\simeq \SU(2)$ holonomy.  Since this solution does not involve
$F_{(4)}$, its generalized holonomy is similarly $\SU(2)$
\begin{equation}
\Hol_{\rm MK}=\SU(2).
\end{equation}
%

\section{Some $n=8$ examples}

Having looked at the basic objects of M-theory, we now turn to
intersecting configurations preserving fewer supersymmetries
\cite{Papadopoulostownsend,Tseytlinharm,Gauntlett}.  While
large classes of intersecting brane solutions and configurations
involving to branes at angles have been constructed, we will only examine
some of the simple cases of orthogonal intersections yielding $n=8$.

\subsection{Branes with a KK-monopole}

It has often been noted that the basic supergravity $p$-brane solutions are
not restricted to having only flat Euclidean transverse spaces.  This
indicates, in particular, that the M5 and M2 solutions of (\ref{eq:m5soln})
and (\ref{eq:m2soln}) demand only that the transverse space spanned by
$\{\vec y\,\}$ is Ricci flat.  Of course, this Ricci flat manifold must
still be supersymmetric in order to preserve some fraction of supersymmetry.

A simple example would be to replace $\E^4$ with a Taub-NUT configuration
in four of the transverse directions to the brane.  For the M5 case, the
resulting M5/MK solution has the form \cite{Gauntlett:1997pk}
\begin{equation}
ds^2=H_5^{-1/3}dx_\mu^2+H_5^{2/3}[dy^2+H_6(dr^2+r^2d\Omega_2^2)+H_6^{-1}
(dz-q_6\cos\theta\,d\phi)^2].
\end{equation}
Here, the M5-brane is delocalized along the $y$ direction, so the
harmonic functions have the form $H_5=1+q_5/r$ and $H_6=1+q_6/r$.
This represents the lifting of a NS5/D6 configuration to eleven dimensions.

Noting that four of the five transverse directions is replaced by a
Taub-NUT space, the corresponding Riemannian holonomy is contained in the
$\SO(5)$ tangent space group in the sense of $\SU(2)\subset\SO(4)\subset
\SO(5)$.  The embedding of the self-dual connection in $\SO(4)$ leads to
explicit $\SU(2)$ generators $T^{(\rm MK)}_{ab}=-\fft{i}2\Gamma_{\bar a
\bar b} P_K^+$ where $P_K^\pm=\fft12(1\pm\Gamma^{1234})$ and
$a,b,\ldots=1,\ldots,4$.  On the other hand, as shown in
section~\ref{sec:m5sec}, the $SO(5)_+$ generalized holonomy of M5 in the
transverse directions involve the $P_5^+$ projection, and is generated by
$T^{(\rm M5)}_{ij}=-\fft{i}2\Gamma_{\bar i\bar j}P_5^+$, where
$i,j,\ldots=1,\ldots\,5$.  As a result, the transverse holonomy of this
M5/MK configuration arises as the closure of $T_{ij}^{(\rm M5)}$ and
$T_{ab}^{(\rm MK)}$.

Since $T_{ab}^{(\rm MK)}$ is comprised of Dirac matrices entirely in
the transverse directions, we may perform a trivial decomposition
\begin{equation}
T_{ab}^{(\rm MK)}=-\ft{i}2\Gamma_{\bar a\bar b}P_K^+P_5^+
-\ft{i}2\Gamma_{\bar a\bar b}P_K^+P_5^-
\end{equation}
Because the first term is already contained entirely in $T_{ij}^{(\rm M5)}$,
the resulting algebra is equally well generated by the mutually commuting set
\begin{equation}
T_{ij}^{(\rm M5)}=-\ft{i}2\Gamma_{\bar i\bar j}P_5^+,\qquad
\widetilde T_{ab}^{(\rm MK)}=-\ft{i}2\Gamma_{\bar a\bar b}P_K^+P_5^-.
\end{equation}
This indicates that the transverse holonomy is simply $\SO(5)_+\times\SU(2)_-$
where $\pm$ refers to the embedding inside the $\hat D$ structure group
$\SO(5)_+\times\SO(5)_-$ for a 6/5 split.
The additional M5 mixed commutator generators
$\{K_\mu,K_{\mu\,i},K_{\mu\,ij}\}$ now transform under both $\SO(5)_+$
and $\SU(2)_-$.  Working out the weights of these generators under $\SO(5)_+
\times\SU(2)_-$ demonstrates that the generalized holonomy of this M5/MK
configuration is
\begin{equation}
\Hol_{\rm M5/MK}=[\SO(5)_+\times\SU(2)_-]\ltimes6\R^{2(4,1)+(4,2)}.
\end{equation}

For the M2-brane, the eight-dimensional transverse space may be given a
hyper-K\"ahler metric \cite{Gauntlett:1997pk}, which is generically of
holonomy $\Sp(2)$.  However, we only consider the product of two
independent Taub-NUT spaces, with holonomy $\Sp(1)\times \Sp(1)$.
Provided both are oriented properly with the M2, this yields a single
additional halving of the supersymmetries, leading to $n=8$.  The
transverse holonomy of this solution corresponds to
the embedding $\SO(8)\times\SU(2)\times\SU(2)\subset\SO(8)\times\SO(4)
\times\SO(4)\subset\SO(8)\times\SO(8)\subset\SO(16)$, where $\SO(16)$
is the $\hat D$ structure group corresponding to a 3/8 split.  The
complete generalized holonomy group is
\begin{equation}
\Hol_{\rm M2/MK/MK}=[\SO(8)\times\SU(2)\times\SU(2)\ltimes3\R^{(8_s,2,2)}]
\ltimes6\R^{2(8_s,1,1)}.
\end{equation}
With only a single Taub-NUT space, the generalized holonomy is instead
\begin{equation}
\Hol_{\rm M2/MK}=[\SO(8)\times\SU(2)\ltimes3\R^{2(8_s,2)}]\ltimes
6\R^{2(8_s,1,1)}.
\end{equation}

\subsection{Branes with a wave}

For solutions with an extended longitudinal space, it is possible to
turn on a wave in a null direction along the brane.  We consider the
M2/MW and M5/MW combinations, both of which preserve a quarter of the
original supersymmetries.  For the M2/MW combination, the supergravity
solution is given by \cite{Tseytlinharm}
\begin{eqnarray}
&&ds^2=H_2^{-2/3}(2dx^+dx^-+K\,dx^{+\,2}+dz^2)+H_2^{1/3}d\vec y\,^2,
\nonumber\\
&&F_{+-zi}=\partial_i\fft1{H_2}.
\end{eqnarray}
Here, both $K$ and $H_2$ are harmonic on the eight-dimensional overall
transverse space; the wave is delocalized along the $z$ direction.

If $H_2$ is turned off, the solution reverts to the MW solution of
(\ref{eq:mwsoln}), however with dependence on only eight of the nine
directions transverse to the wave.  The resulting holonomy would be
$\R^8$.  Combining this with the M2 generalized holonomy,
(\ref{eq:m2hol}), must yield a larger group that is nevertheless
contained in $\SL(24,\R)\ltimes8\R^{24}$.

To see this explicitly, we first note that the generalized covariant
derivative has the form
\begin{eqnarray}
\D_+&=&\partial_++\ft14H_2^{-1/2}\partial_iK\Gamma_-\Gamma_{\bar i}
-\ft16H_2^{-3/2}\partial_i H_2(\Gamma_++\ft12K\Gamma_-)\Gamma_{\bar i}P_2^+,
\nonumber\\
\D_-&=&\partial_--\ft16H_2^{-3/2}\partial_i H_2\Gamma_-\Gamma_{\bar i}P_2^+,
\nonumber\\
\D_z&=&\partial_z-\ft16H_2^{-3/2}\partial_i H_2\Gamma_{\bar z}\Gamma_{\bar i}
P_2^+,
\nonumber\\
\D_i&=&\partial_i+\ft1{12}(\Gamma_{\bar i}{}^{\bar j}-2\delta_i^j)P_2^+
\partial_j\ln H_2-\ft16\partial_i\ln H_2,
\end{eqnarray}
where all Dirac matrices are written with frame indices.  As usual, the M2
projection is defined by $P_2^\pm=\fft12(1\pm\Gamma^{(2)})$, where
$\Gamma^{(2)}=\Gamma^{+-\bar z}$.

Taking commutators of the above covariant derivatives, it is clear that
the generalized holonomy algebra is formed by the closure of the MW
algebra, generated by $\Gamma_-\Gamma_{\bar i}$, and the M2 algebra,
generated by $-\fft{i}2\Gamma_{\bar i\bar j}P_2^+$ and
$\Gamma_{\bar\mu\bar i}P_2^+$ where $\mu$ denotes one of the longitudinal
coordinates, $+$, $-$ or $z$.  To be explicit, we may use the Dirac matrix
decomposition given by (\ref{eq:m2dirac}).  The light-cone Dirac matrices
are then given by
\begin{equation}
\Gamma^+=\ft1{\sqrt{2}}(-\Gamma^0+\Gamma^1),\qquad
\Gamma^-=\ft1{\sqrt{2}}(-\Gamma^0-\Gamma^1),
\end{equation}
so that $\Gamma^{(2)}=\Gamma^{012}=1\times1\times\sigma^3$, and the
M2 projection has the identical form as (\ref{eq:m2proj}).  Killing spinors
for the wave solution are projected according to $\Gamma_-\epsilon=0$,
or equivalently $\Gamma^+\epsilon=0$.  In terms of $0$ and $1$ components,
this corresponds to the condition $P_L^+\epsilon=0$ where the wave
projection is given by $P_L^\pm=\ft12(1\pm\Gamma^{01})$.

The M2/MW Killing spinors satisfy the simultaneous conditions $P_2^+\epsilon=0$
and $P_L^+\epsilon=0$, where $P_2^+$ is given in (\ref{eq:m2proj}) and
\begin{equation}
P_L^+=1\times\pmatrix{1&0\cr0&0}\times1.
\end{equation}
Combining the last two elements in the three-term direct product, the
projections may be explicitly written as
\begin{equation}
P_2^+=1\times\pmatrix{1\cr&0\cr&&1\cr&&&0},\qquad
P_L^+=1\times\pmatrix{1\cr&1\cr&&0\cr&&&0}.
\end{equation}
As a result, Killing spinors are given by
\begin{equation}
\epsilon=\eta\times\pmatrix{0\cr0\cr0\cr1},
\end{equation}
and a typical generator of the generalized holonomy group must have the form
\begin{equation}
T=\pmatrix{\ddots&&\rddots&0\cr&\SL(24,\R)&&0\cr\rddots&&\ddots&0\cr
\cdots&\R(8,24)&\cdots&0}.
\end{equation}

In this $4\times4$ matrix notation, the M2 holonomy generators of
(\ref{eq:m2holgen}) may be written as
\begin{equation}
T_{\rm M2}=\pmatrix{\SO(8)_+&0&0&0\cr A&0&B&0\cr0&0&\SO(8)_+&0\cr C&0&A&0},
\end{equation}
where the single $\SO(8)_+$ transverse holonomy simultaneously transforms
the first and third entries of the four-component vector.  Here, $A$, $B$
and $C$ are independent $\GL(8,\R)$ matrices.  In addition, the $\R^8$
holonomy of the wave (delocalized along $z$) is generated by
\begin{equation}
T_{\rm MW}=\pmatrix{0&0&0&0\cr0&0&0&0\cr0&b^01+ib^a\gamma^a&0&0\cr
b^01-ib^a\gamma^a&0&0&0},
\end{equation}
where $\{b^0,b^a\}$ is an eight-component vector.  Closing the algebra
generated by $T_{\rm M2}$ and $T_{\rm MW}$ results in the M2/MW generators
\begin{equation}
T_{\rm M2/MW}=\pmatrix{\SO(8)&0&0&0\cr\R(8,16)&\multispan2\hfil$\SL(16,\R)$&0\cr
\vdots&\multispan2\hfil~~\vdots\hfil&0\cr\cdots&\R(8,24)&\cdots&0}.
\end{equation}
Thus the corresponding generalized holonomy group is
\begin{equation}
\Hol_{\rm M2/MW}=[\SO(8)\times\SL(16,\R)\ltimes \R^{(8,16)}]\ltimes8\R^{(8,1)
+(1,16)}.
\end{equation}

The generalized holonomy analysis for the M5/MW solution \cite{Tseytlinharm}
\begin{eqnarray}
&&ds^2=H_5^{-1/3}(2dx^+dx^-+K\,dx^{+\,2}+d\vec z_4{}^2)+H_5^{2/3}d\vec y\,^2,
\nonumber\\
&&F_{ijkl}=\epsilon_{ijklm}\partial_mH_5,
\end{eqnarray}
is similar.  Here the functions $H_5$ and $K$ are harmonic on the
five-dimensional overall transverse space.  This corresponds to a
superposition of a M5-brane with a delocalized wave, where the latter
has $\R^5$ holonomy.  Closing the holonomy algebra over the M5 and MW
generators yields the generalized holonomy
\begin{equation}
\Hol_{\rm M5/MW}=[\SO(5)\times\SU^*(8)\ltimes4\R^{(4,8)}]\ltimes8\R^{2(4,1)
+2(1,8)}.
\end{equation}
Note that $\SU^*(8)\simeq\SL(4,\Hq)$, and the latter is built out of multiple
copies of the five-dimensional real Clifford algebra $\Cl(5,0)_+
\simeq\GL(2,\Hq)$.

\subsection{Other examples}

Additional pure geometry backgrounds may be constructed by combining a
wave with a Taub-NUT space.  An $n=8$ example is given by
\cite{Tseytlinharm,Bergshoeffderoo}
\begin{equation}
ds^2=dx^+dx^-+K\,dx^{+\,2}+d\vec y_5{}^2+H_6(dr^2+r^2d\Omega_2^2)
+H_6^{-1}(dz-q_6\cos\theta\,d\phi)^2,
\end{equation}
where $K=q_0/r+q_y/y^3$ and $H_6=1+q_6/r$.  Since the transverse space
is a direct product of $\E^5$ with Taub-NUT, the generalized holonomy
has the direct product form
\begin{equation}
\Hol_{\rm MW/MK}=\R^5\times(\SU(2)\ltimes\R^{2(2)}).
\end{equation}

Finally, there are numerous examples of overlapping or intersecting brane
configurations involving multiple M2 and/or M5 branes.  Various fractions
of supersymmetry may be preserved by placing branes at appropriate angles.
Here, we only consider the orthogonal intersection of M2 and M5 on a
string, given by \cite{Tseytlinharm,Gauntlett}
\begin{eqnarray}
&&ds^2=H_2^{-2/3}H_5^{-1/3}dx_\mu^2+H_2^{1/3}H_5^{-1/3}d\vec w_4{}^2
+H_2^{-2/3}H_5^{2/3}dz^2+H_2^{1/3}H_5^{2/3}d\vec y_4{}^2,\nonumber\\
&&F_{\mu\nu zi}=\epsilon_{\mu\nu}\partial_i\fft1{H_2},\qquad
F_{ijk z}=\epsilon_{ijkl}\partial_lH_5.
\end{eqnarray}
The full holonomy algebra is obtained by the closure of the M5 and M2
holonomies, given by (\ref{eq:m5hol}) and (\ref{eq:m2hol}), respectively.
A slight complication arises in that the individual generators work on
different relative transverse directions for the M5 and M2 branes.  By
taking the non-compact generators of one of the branes ({\it e.g.} the
$12\R^{2(8_s)}$ for the M2) and commuting with the transverse holonomy
generators of the other (in this case the $SO(5)_+$ for the M5) we end
up filling up all of $\SL(24,\R)$.  As a result, we find that the M2/M5
generalized holonomy fills all of the maximally allowed case for $n=8$,
namely
\begin{equation}
\Hol_{\rm M2/M5}=\SL(24,\R)\ltimes8\R^{24}.
\end{equation}
%

\section{Waves and supernumerary Killing spinors}
\label{sec:ppwave}

In this section, we consider waves with non-vanishing $F_{(4)}$.  For
a pp-wave with covariantly constant null Killing vector
$\partial/\partial x^-$, the metric and four-form take the form
\begin{eqnarray}
ds^2&=&2\,dx^+dx^-+K\,dx^{+\,2}+d\vec y\,^2,\nonumber\\
F_{(4)}&=&\mu\,dx^+ \wedge \Phi_{(3)},
\end{eqnarray}
where $\mu$ is a nonzero constant and $\Phi_{(3)}$ is a harmonic three-form on
the transverse space.  In general, the function $K$ depends on both $x^+$ and
$\vec y$, while for plane waves, it has the quadratic form
$K=K_{ij}(x^+)y^iy^j$.

The metric is identical to that of (\ref{eq:mwsoln}), which was considered
previously in the pure geometry case.  Thus the generalized covariant
derivative is given by
\begin{eqnarray}
\D_+&=&D_+ -\ft{i}{12}\mu(1+\Gamma_-\Gamma_+)W,\nonumber\\
\D_-&=&\partial_-,\nonumber\\
\D_i&=&\partial_i + \ft{i}{24}\mu\Gamma_-(\Gamma_iW+3W\Gamma_i),
\end{eqnarray}
where $W=\fft{i}{3!}\Phi_{ijk}\Gamma_{ijk}$, and the
gravitational covariant derivative $D_+$ is given in (\ref{eq:ppcder}).
With non-vanishing $F_{(4)}$, the integrability condition of (\ref{eq:commw})
is modified to become $M_{+i}\,\epsilon=0$ where
\begin{equation}
M_{+i}=-\ft14[\partial_i\partial_jK\Gamma_j+\ft{\mu^2}{72}(6W\Gamma_iW+
9W^2\Gamma_i+\Gamma_iW^2)]\Gamma_-\equiv-\ft14X_i\Gamma_-.
\end{equation}
Note that this integrability condition acting on a spinor $\epsilon$ is
in exact agreement with the first order Killing spinor conditions for the
pp-wave background \cite{Cvetic:2002si,Gauntletthull2}.  In particular,
since $(\Gamma_-)^2=0$, half of the original supersymmetries ($n=16$) are
always preserved by spinors satisfying $\Gamma_-\epsilon=0$.  On the
other hand, extra supersymmetries (denoted supernumerary supersymmetries
in \cite{Cvetic:2002si}) arise whenever $X_i$ has zero eigenvalues.

If we identify the generalized holonomy generators as $T^i=X_i\Gamma_-$,
and furthermore note that $\{X_i,\Gamma_-\}=0$, it is easy to see that
$[T^i,T^j]=0$.  Hence the nine generators fill out at most $\R^9$.  But
the generalized holonomy may be smaller if any of the generators are
either degenerate or trivial.  In particular, the generalized holonomy
group must be trivial for the maximally supersymmetric Hpp-wave
\cite{Kowalski-Glikman:wv,Hull,Figueroa-O'Farrill:2001nz}.

To investigate the generalized holonomy for plane waves with exotic
fractions of supersymmetry, consider the ansatz
\cite{Cvetic:2002si,Gauntletthull2}
\begin{eqnarray}
\Phi_{(3)}&=&m_1\,dy_{129}+m_2\,dy_{349}+m_3\,dy_{569}+m_4\,dy_{789},
\nonumber\\
K&=&1-\sum_i{\mu_i^2y_i^2},
\end{eqnarray}
where $dy_{ijk}=dy_i\wedge dy_j\wedge dy_k$, and the equations of motion
demand $\sum\mu_i^2=\fft1{12}\mu^2\Phi^2$.  The $\mu_i$ must be chosen
appropriately in order to preserve supersymmetry \cite{Gauntletthull2}.
Since the direction
$i=9$ is singled out, the result is somewhat asymmetrical, with
$\mu_9^2=\fft19\mu^2(m_1+m_2+m_3+m_4)^2$, while $\mu_1^2=\mu_2^2
=\fft1{36}(2m_1-m_2-m_3-m_4)^2$ with similar expressions for
$\mu_3^2,\ldots,\mu_8^2$ (where the factor of 2 is permuted).  In
this case, we find
\begin{eqnarray}
X_{1,2}&=&-\ft1{18}\mu^2\Gamma_{1,2}[(2m_1-m_2-m_3-m_4)^2
-(2m_1-m_2\Gamma^{1234}-m_3\Gamma^{1256}-m_4\Gamma^{1278})^2],\nonumber\\
&\vdots&\cr
X_9&=&-\ft29\mu^2\Gamma_9[(m_1+m_2+m_3+m_4)^2-(m_1+m_2\Gamma^{1234}
+m_3\Gamma^{1256}+m_4\Gamma^{1278})^2].
\label{eq:xieqn}
\end{eqnarray}

The choice of setting all $m_i$ to zero trivially recovers the Minkowski
vacuum, with $n=32$.  On the other hand, even when exactly one of the $m_i$
is nonzero, all the $X_i$ still vanish.  This case corresponds to the
Hpp-wave, which preserves all supersymmetries ($n=32$), and which has
trivial generalized holonomy
\begin{equation}
\Hol_{\rm Hpp}=\{1\}.
\end{equation}
We may see that with each additional non-vanishing $m_i$ turn on, the $X_i$
take on the form of multiple commuting projections, with the projections
built from $\Gamma^{1234}$, $\Gamma^{1256}$ and finally $\Gamma^{1278}$.
Hence this appropriate connection between the $\mu_i$ (metric) and $m_i$
(four-form) constants allows the addition of 2, 4 or 8 supernumerary
supersymmetries.  A slightly different ansatz for $\Phi_{(3)}$ also allows
for 6 extra supersymmetries.  Thus in this manner we obtain plane waves
with $n=18,20,22$ and $24$ \cite{Gauntletthull2}.

For all these cases with $n<32$, none of the $X_i$ in (\ref{eq:xieqn})
vanish.  Since each individual $X_i$ is obtained by multiplying $\Gamma_i$
by a suitable projector, they are all linearly independent.  Hence the
generalized holonomy remains $\R^9$, regardless of the actual number of
supersymmetries.  Furthermore, even if the $\mu_i$ and $m_i$ were not
chosen appropriately (so that there are no extra supersymmetries), the
plane wave would still preserve the original $n=16$.  Hence
\begin{equation}
\Hol_{\rm pp}=\R^9\qquad (n=16, 18, 20, 22, 24).
\end{equation}
The $n=16$ case is essentially that of the MW found before.

As seen in (\ref{eq:xieqn}), the projections which are responsible for
the exotic fractions of supersymmetries are hidden inside the $X_i$.
Without a detailed examination of the generators $T^i=X_i\Gamma_-$, we
cannot tell how many extra supersymmetries there are simply by looking
at the generalized holonomy group itself.

\section{Discussion}

As we have seen in the previous three sections, the generalized holonomy of
M-theory solutions takes on a variety of guises.  Our results are summarized
in table~\ref{tab:1}.  We make note of two features exhibited by these
solutions.  Firstly, it is clear that many generalized holonomy groups
give rise to the same number $n$ of supersymmetries.  This is a consequence
of the fact that while $\Hol$ must satisfy the condition (\ref{n}), there
are nevertheless many possible subgroups of $\SL(32-n,\R)\ltimes
n\R^{(32-n)}$ allowed by generalized holonomy.  Secondly, as demonstrated
by the plane wave solutions, knowledge of $\Hol$ by itself is insufficient
for determining $n$; here $\Hol=\R^9$, while $n$ may be any even integer
between 16 and 26.

\begin{table}[t]
\begin{tabular}{lll}
$n$&Background&Generalized holonomy\\
\hline
32&$\E^{1,10}$, AdS$_7\times S^4$\qquad\qquad&$\{1\}$\\
&AdS$_4\times S^7$, Hpp\\
\hline
18,\ldots,26&plane waves&$\R^9$\\
\hline
16&M5&$\SO(5)\ltimes6\R^{4(4)}$\\
16&M2&$\SO(8)\ltimes12\R^{2(8_s)}$\\
16&MW&$\R^9$\\
16&MK&$\SU(2)$\\
\hline
8&M5/MK&$[\SO(5)\times\SU(2)]\ltimes6\R^{2(4,1)+(4,2)}$\\
8&M2/MK/MK&$[\SO(8)\times\SU(2)\times\SU(2)\ltimes3\R^{(8_s,2,2)}]
\ltimes6\R^{2(8_s,1,1)}$\\
8&M2/MK&$[\SO(8)\times\SU(2)\ltimes3\R^{2(8_s,2)}]\ltimes6\R^{2(8_s,1,1)}$\\
8&M2/MW&$[\SO(8)\times\SL(16,\R)\ltimes \R^{(8,16)}]\ltimes8\R^{(8,1)
+(1,16)}$\\
8&M5/MW&$[\SO(5)\times\SU^*(8)\ltimes4\R^{(4,8)}]\ltimes8\R^{2(4,1)+2(1,8)}$\\
8&MW/MK&$\R^5\times(\SU(2)\ltimes\R^{2(2)})$\\
8&M2/M5&$\SL(24,\R)\ltimes8\R^{24}$\\
\hline
\end{tabular}
\caption{Generalized holonomies of the objects investigated in the text.
For $n=16$, we have $\Hol\subseteq\SL(16,\R)\ltimes16\R^{16}$, while
for $n=8$, it is instead $\Hol\subseteq\SL(24,\R)\ltimes8\R^{24}$.}
\label{tab:1}
\end{table}

What this indicates is that, at least for counting supersymmetries, it
is important to understand the embedding of $\Hol$ in $\cal G$.  In
contrast to the Riemannian case, different embeddings of $\Hol$ yield
different possible values of $n$.  Although this appears to pose a
difficulty in applying the concept of generalized holonomy towards
classifying supergravity solutions, it may be possible that a better
understanding of the representations of non-compact groups will nevertheless
allow progress to be achieved in this direction.

While the full generalized holonomy involves several factors, the transverse
(or $\hat D$) holonomy is often simpler, {\it e.g.} $SO(5)$ for the M5 and
$SO(8)$ for the M2.  The results summarized in table~\ref{tab:1} are
suggestive that the maximal compact subgroup of $\Hol$, which must be
contained in $\SL(32-n,\R)$, is often sufficient to determine the number of
surviving supersymmetries.  For example, the M2/MK/MK solution may be
regarded as a 3/8 split, with a hyper-K\"ahler eight-dimensional
transverse space.  In this case, the $\hat D$ structure group is $\SO(16)$,
and the 32-component spinor decomposes under $\SO(32)\supset
\SO(16)\supset\SO(8)\times\SU(2)\times\SU(2)$ as $32\to2(16)\to
2(8,1,1)+2(1,2,2)+8(1,1,1)$ yielding eight singlets.  Similarly, for the
M5/MW intersection, we consider a 2/9 split, with the wave running along
the two-dimensional longitudinal space.  Since the $\hat D$ structure
group is $\SO(16)\times \SO(16)$ and the maximal compact subgroup of
$\SU^*(8)$ is ${\rm USp}(8)$, we obtain the decomposition
$32\to(16,1)+(1,16)\to4(4,1)+(1,8)+8(1,1)$ under $\SO(32)\supset
\SO(16)\times\SO(16)\supset\SO(5)\times{\rm USp}(8)$.  This again yields
$n=8$.  Note, however, that this analysis fails for the plane waves, as
$\R^9$ has no compact subgroups.

A different approach to supersymmetric vacua in M-theory is 
through the technique of $G$-structures \cite{Gauntlett:2002fz}. Hull 
\cite{Hull:2003mf} has suggested that $G$-structures may be better 
suited to finding supersymmetric solutions whereas generalized holonomy may be better 
suited to classifying them. In any event, it would be useful to 
establish a dictionary for translating one technique into the other. 

Ultimately, one would hope to achieve a complete classification of
vacua for the full M-theory.  
In this regard, one must at least include the effects of M-theoretic corrections to the 
supergravity field equations and Killing spinor equations and perhaps 
even go beyond the geometric picture altogether. It seems likely, 
however, that counting supersymmetries by the number of singlets 
appearing in the decomposition $32$ of $\SL(32,\R)$ under 
$\Hol \subset \SL(32,\R)$ will continue to be valid.

\section*{Acknowledgments}

We have enjoyed useful conversations with Lilia Anguelova,
Jianxin Lu, Juan Maldacena, Malcolm Perry, Hisham Sati and Dimitrios
Tsimpis.


\end{document}